\documentclass[aps,pra,showpacs,twocolumn,floatfix,nofootinbib,superscriptaddress]{revtex4}
\usepackage{graphicx} 
\usepackage{amsmath}
\usepackage{bm}
\usepackage{dcolumn}
\begin{document}

\title{
Precision determination of electroweak coupling from atomic parity violation and
implications for particle physics
}

\author{S. G. Porsev}
\affiliation{Department of Physics, University of Nevada, Reno,
Nevada 89557, USA}
\affiliation{Petersburg Nuclear Physics Institute, Gatchina
188300, Russia}
\author{K. Beloy}
\affiliation{Department of Physics, University of Nevada, Reno,
Nevada 89557, USA}

\author{A. Derevianko}
\affiliation{Department of Physics, University of Nevada, Reno,
Nevada 89557, USA}


\date{\today}
\begin{abstract}
We carry out high-precision calculation of parity violation in cesium atom, reducing theoretical uncertainty by a factor of two compared
to previous evaluations. We combine previous measurements with our calculations and extract the weak charge of the $^{133}$Cs nucleus,
$Q_W = -73.16(29)_\mathrm{exp}(20)_\mathrm{th}$. The result is in agreement with the Standard Model (SM) of elementary particles.
This is the most accurate to-date test of the low-energy electroweak sector of the SM.
In combination with the results of high-energy collider experiments, we confirm the energy-dependence (or ``running'') of the electroweak force over an energy range spanning four orders of magnitude (from $\sim 10$ MeV  to  $\sim 100$ GeV). Additionally, our result places constraints on a variety of new physics scenarios beyond the SM. In particular, we increase the lower limit on the masses of extra $Z$-bosons predicted by models of grand unification and string theories.
\end{abstract}


\maketitle

Atomic parity violation  places powerful constraints on
new physics beyond the Standard Model (SM) of elementary particles~\cite{MarRos90,Ram99}.
The measurements are interpreted in terms of the nuclear weak
charge $Q_W$, quantifying the strength of the electroweak coupling between atomic
electrons and quarks of the nucleus. Here we report the most accurate to-date
determination of this coupling strength by combining previous measurements~\cite{WooBenCho97,BenWie99} with
our high-precision calculations in cesium atom. The result,
$Q_W(^{133}\mathrm{Cs}) = -73.16(29)_\mathrm{exp}(20)_\mathrm{th}$, is in a perfect agreement
with the  prediction of the SM. In combination with the results of high-energy collider experiments, our work confirms the predicted energy-dependence (or ``running'') of the electroweak interaction over an energy range spanning four orders of magnitude (from $\sim 10$ MeV  to  $\sim 100$ GeV).
%
The attained precision is important  for probing ``new physics''.
As an illustration, we constrain the mass of
the so-far elusive particle -- the extra $Z$ boson ($Z'$). $Z'$ are hypothesized to be carriers of the ``fifth force''
of Nature, and they are abundant in models of grand unification and string theories~\cite{ZprimeRefs}.
In particular,
SO(10) unification predicts a $Z'$-boson denoted as $Z'_\chi$. A direct search at Tevatron collider ~\cite{AalAbuAde07} yielded
$M_{Z'_\chi} > 0.82 \, \mathrm{TeV}/c^2$. Our precision result implies
a more stringent bound,  $M_{Z'_\chi} > 1.3 \, \mathrm{TeV}/c^2$. If $Z'$ is discovered at the Large Hadron Collider (LHC), where the mass scale reach is somewhat higher, our result would help in exacting  $Z'$ properties.

Historically, atomic parity violation helped in establishing the validity of the SM~\cite{Khr91,BouBou97,GinFla04}.
While a  number of experiments have been carried out,
the most accurate  measurement is due to Wieman and collaborators~\cite{WooBenCho97}.
They determined a ratio of the parity nonconserving (PNC) amplitude, $E_\mathrm{PNC}$, to the vector transition polarizability, $\beta$,
$E_\mathrm{PNC}/\beta=1.5935(56) \, \mathrm{mV/cm}$,
on the parity-forbidden $6S_{1/2} \rightarrow 7S_{1/2}$
transition in atomic Cs. 

The measurement, however, does not directly translate into
an electroweak observable of the same accuracy, as
the interpretation of the experiment requires input from atomic theory. The theory links  $Q_W$ to the measured signal.
In computations, $Q_W$ is treated as a parameter, and by combining computed $E_\mathrm{PNC}$ with
measurements, the value of $Q_W$ is derived. The inferred $Q_W$  is compared with the predicted SM value, either
revealing or constraining new physics. So far the atomic-theory error has been a limiting factor in this interpretation. Here we
report reducing this error, leading to an improved test of the SM.

The PNC amplitude for the $6S_{1/2} \rightarrow 7S_{1/2}$ transition
in Cs may be evaluated as
\begin{eqnarray}
\lefteqn{E_\mathrm{PNC} = \sum_{n}
\frac{\langle 7S_{1/2}|D_z|nP_{1/2}\rangle  \langle nP_{1/2} |H_{W}|6S_{1/2}\rangle
}{E_{6S_{1/2}}-E_{nP_{1/2}}}  } \nonumber \\  &+ &
\sum_{n}
\frac{\langle 7S_{1/2}|H_{W}|nP_{1/2}\rangle  \langle nP_{1/2} |D_z|6S_{1/2}\rangle
}{E_{7S_{1/2}}-E_{nP_{1/2}}}
\, .
\label{Eq:EPNC}
\end{eqnarray}
Here $D$ and $H_{\rm W}$ are electric-dipole and weak interaction operators,
and $E_{i}$ are atomic energy levels. In the electronic sector, the effective weak interaction averaged over quarks reads
$
 H_{\rm W} = -\frac{G_F}{\sqrt{8}} \, Q_W \,  \gamma_5 \,
 \rho ({\bf r}) $,
where $G_F$ is the Fermi constant, $\gamma_5$ is the Dirac matrix, and
$\rho ({\bf r})$ is the  neutron-density distribution. $^{133}$Cs nucleus has $Z=55$ protons and $N=78$ neutrons.
The value of  $Q_W$ is given approximately by $-N$.

\begin{figure}[h]
\begin{center}
\includegraphics*[scale=0.35]{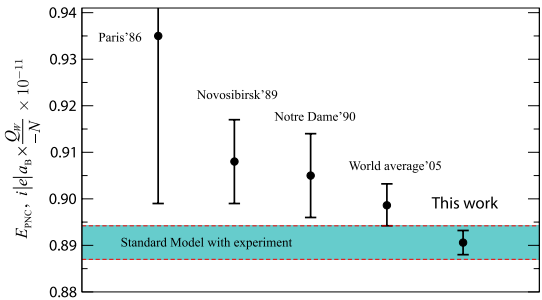}
\caption{(Color online)  Progress in evaluating the PNC amplitude.
Points marked Paris '86, Novosibirsk '89, Notre Dame '90
correspond to Refs.~\cite{BouPik86}, \cite{DzuFlaSus89}, and~\cite{BluJohSap90}.
Point ``World average '05'' is due to efforts of several groups~\cite{Der00,Der02,theorPNCextraNewBefore2005,MilSusTer02,ShaPacTup05}
on sub-1\% Breit, QED, and neutron-skin corrections reviewed in
Ref.~\cite{DerPor07}. The strip corresponds to a combination of the Standard Model $Q_W$ with
measurements~\cite{WooBenCho97,BenWie99}. The edges of the strip correspond to $\pm \sigma$ of the measurement.
Here we express $E_\mathrm{PNC}$ in
conventional units of $ i |e| a_{\mathrm B} \left( -{Q_W}/{N} \right) \times 10^{-11}$,
where $e$ is the elementary charge and $a_{\mathrm B}$ is the Bohr radius. These units
factor out a ratio of $Q_W$ to its approximate value, $-N$.
\label{Fig:compEPNC}}
\end{center}
\end{figure}

Interpretation of the PNC measurements requires evaluating Eq.(\ref{Eq:EPNC}).
Although the underlying theory of quantum electrodynamics (QED) is well established, the atomic many-body problem is intractable.
Reaching theoretical accuracy
equal to or better than the experimental accuracy of 0.35\%
has been a challenging task (see Fig.~\ref{Fig:compEPNC}).
An important  1\% accuracy  milestone has been reached by the Novosibirsk~\cite{DzuFlaSus89}
and Notre Dame~\cite{BluJohSap90} groups in the late 1980s.
More recently, several groups have contributed to understanding sub-1\% corrections, primarily due to
the Breit (magnetic) interaction and radiative QED processes~\cite{Der00,Der02,theorPNCextraNewBefore2005,MilSusTer02,ShaPacTup05} (reviewed in \cite{DerPor07}). The results of these calculations
are summarized by the ``World average '05'' point of Fig.~\ref{Fig:compEPNC}, which has a 0.5\% error bar reflecting this progress.
As of 2005, the sensitivity to new physics has been limited by the accuracy of solving the basic correlation problem.
Here we report an important progress in solving it.

We wish to  evaluate accurately the sum~(\ref{Eq:EPNC}).
 To this end,
we solve the Schrodinger equation $H |\Psi_v\rangle = E_v |\Psi_v\rangle$
and find atomic wave functions and energies.
%
%
Even in classical mechanics, the simpler {\em three}-body problem  cannot be solved in closed form.
For Cs atom, one
solves for a correlated motion of 55 electrons. The problem is simplified by the fact that
this atom has one loosely-bound valence electron $v$ outside a stiff closed-shell core.
Because of that, the problem can be efficiently treated within the many-body perturbation theory~\cite{LinMor86}.
In this treatment, the exact many-body state $|\Psi_v\rangle$ which
stems from the approximate (Dirac-Fock) state $|\Psi_v^{(0)} \rangle$ is parameterized as
$|\Psi_v\rangle = \Omega\, |\Psi_v^{(0)}\rangle \, ,$
where the many-body operator  $\Omega$ is yet to be found.
It is expanded into a hierarchy of
single-, double-, triple-, and higher-rank $n$-fold excitations.
For example, double excitations (or simply doubles)  result from a simultaneous scattering of two core electrons by their mutual
Coulomb repulsion.

Notice that for the 55 electrons of Cs this treatment would be exact by including
55-fold excitations. However, manipulating such wave functions is impractical: for a basis set of 100 orbitals,
one would require more than $100^{55}$ memory units. This number exceeds the estimated number of atoms in the Universe.
Fortunately, contributions of high-rank excitations are strongly suppressed.
Previous many-body calculations\cite{DzuFlaGin02,BluJohSap90} in Cs stored only single and double excitations.
Already at this level the attained accuracy for the atomic properties of Cs was at the level of 1\% or better.
To systematically improve the accuracy,
here we take advantage of modern computing resources and additionally
store and manipulate {\em triple} excitations. This is a substantial step. For example,  previous calculations~\cite{BluJohSap90}
used less than 100 Mb of storage, whereas our calculations required 100 Gb; this is a factor of 1,000 increase in computational
complexity.

Our specific scheme~\cite{DerPor05,PorDer06Na,DerPor07,DerPorBel08} of solving the atomic many-body problem is rooted in the coupled-cluster method~\cite{LinMor86}.
We refer to our approximation as the CCSDvT scheme (Coupled-Cluster approximation including Singles, Doubles, and valence Triples).
Details will be provided elsewhere.
The solution is {\em ab initio} relativistic, as near the Cs nucleus (where the weak interaction occurs)
the electrons move with speeds approaching the speed of light.
To minimize human errors, the CCSDvT code was developed independently by at least two persons.
Complex derivations and coding were aided by symbolic algebra tools. 
An important proof of the code was made by computing properties
of lithium atom~\cite{DerPorBel08}. This atom has 3 electrons, making the CCSDvT approximation
exact.
We found in Ref.~\cite{DerPorBel08} that  experimental data for Li were reproduced numerically with an
accuracy reaching 0.01\%.

Now we proceed to evaluating the PNC amplitude, Eq.~(\ref{Eq:EPNC}), by directly summing over the intermediate $nP_{1/2}$
states~\cite{BluJohSap90}. This implies computing wave functions and energies of the $6S_{1/2}$, $7S_{1/2}$, and $nP_{1/2}$ states, forming matrix elements, and substituting them into Eq.~(\ref{Eq:EPNC}).
We employ a computationally expensive CCSDvT
method only for matrix elements involving $n=6,7,8,9$  (``main'' term) and compute suppressed contributions of
$n\ge10$ and core-excited states  (``tail'' term) with less accurate methods.

Our results for the PNC amplitude are presented in Table~\ref{Tab:EPNC}.
The upper panel of the table lists contributions 
due to the Coulomb interaction of electrons with the nucleus and other electrons. The lower panel summarizes well-established non-Coulomb contributions such as Breit, radiative (QED), and other smaller corrections. Estimated uncertainties are listed in parentheses.

\begin{table}[h]
\caption{ Contributions to the parity violating amplitude $E_\mathrm{PNC}$  for the $6S_{1/2} \rightarrow 7S_{1/2}$ transition in $^{133}$Cs in units of $ i |e| a_{\mathrm B} \left( -\frac{Q_W}{N} \right) \times 10^{-11}$.
\label{Tab:EPNC} }
\begin{center}
\begin{tabular}{lr@{.}l}
\hline \hline
\multicolumn{3}{c}{Coulomb interaction}\\
Main ($n=6-9$)     & 0&8823(18) \\
Tail               & 0&0175(18) \\
\smallskip
Total correlated   & 0&8998(25)   \\
\hline
\multicolumn{3}{c}{Corrections}\\
Breit, Ref.~\cite{Der00}            & -0&0054(5) \\
QED,   Ref.~\cite{ShaPacTup05}       & -0&0024(3) \\
Neutron skin, Ref.~\cite{Der02}     & -0&0017(5) \\
$e-e$ weak interaction,   Ref.~\cite{BluJohSap90}              &  0&0003 \\
\hline
Final & 0&8906(26) \\
\hline \hline
\end{tabular}
\end{center}
\end{table}

\begin{figure}[h]
\begin{center}
\includegraphics*[scale=0.3]{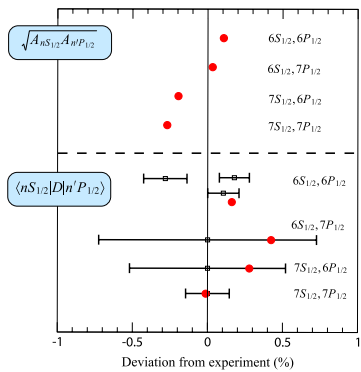}
\caption{(Color online)  Deviations of  computed values (red filled circles) from experimental data (centered at zero).
The upper panel displays combinations of magnetic hyperfine structure constants $\sqrt{A_{nS_{1/2}} A_{n'P_{1/2}}}$
which mimic matrix elements of the weak interaction. For these combinations, experimental error bars are
negligible compared to the theoretical accuracy.
The lower panel exhibits deviations of the computed dipole matrix elements from the most accurate experimental results~\cite{lifetimesCsRefs,VasSavSaf02}.
\label{Fig:comp}}
\end{center}
\end{figure}

We start by assessing the accuracy of the employed CCSDvT approximation.
This determines uncertainty of the ``main'' term contributing 99\% of $E_\mathrm{PNC}$.
Properties of low-energy states have previously been measured, and we quantify theory uncertainties by comparing these
data with our {\em ab initio} results.
For consistency we add QED, Breit, and nuclear-structure corrections to our Coulomb-correlated results.
We find that the experimental energies are reproduced with an accuracy of 0.1-0.3\%. Dipole matrix elements enter the PNC amplitude directly and are derived
from atomic lifetime measurements. Relevant dipoles are compared in the lower panel of Fig.~\ref{Fig:comp}; the CCSDvT values are within the error bars of the experiments.
Finally, since  the hyperfine constants $A$ arise due to interactions of electrons with nuclear magnetic moments,
matrix elements of the weak interaction $\langle nS_{1/2} |H_W| n'P_{1/2} \rangle$ may be tested
by forming the geometric mean $\sqrt{A_{nS_{1/2}} A_{n'P_{1/2}}}$, Ref.~\cite{DzuFlaGin02}.
Deviations of these combinations from experimental data are shown in the upper panel of Fig.~\ref{Fig:comp}. We find that the standard deviation of theoretical values from experiment is 0.2\%.

\begin{figure}[h]
\begin{center}
\includegraphics*[scale=0.3]{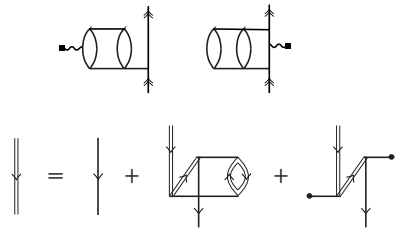}
\caption{ Many-body diagrams responsible for the shift of the PNC amplitude compared to previous calculations.
 Top row: sample direct contributions of valence triples to matrix elements (wavy capped line)~\protect\cite{PorDer06Na}. Bottom row: iterative equation for line dressing of the hole line in expressions for matrix elements~\cite{DerPor05} (similar equation holds for particle lines; exchange diagrams are not shown).
\label{Fig:diags}}
\end{center}
\end{figure}

Overall agreement of theoretical data with experiments indicates that the average accuracy of the CCSDvT
approximation is 0.2\% and we assign an error of 0.2\% to the main term. Additionally, our semi-empirical fitting
to experimental energies modifies the main term by 0.2\%, which is consistent with the above error estimate.
Finally, the ``tail''
was computed using a blend of many-body approximations and we assign a  10\% uncertainty
to this contribution based on the spread of its value in different approximations.  The final result (Table~\ref{Tab:EPNC}) includes
smaller non-Coulomb corrections and its uncertainty was estimated by adding individual uncertainties in quadrature.
Previous calculations~\cite{BluJohSap90,DzuFlaGin02} report values larger by  0.9\% than our 0.27\%-accurate result.
The difference is due to our inclusion of additional  many-body effects, shown in Fig.~\ref{Fig:diags}. Direct contribution of triple excitations to matrix elements accounts for a 0.3\% shift  and dressing of matrix elements for another 0.3\%. The remaining 0.3\% comes
from a consistent removal of QED and Breit corrections from experimental energies during the semi-empirical fit.

With the computed $E_\mathrm{PNC}$
we proceed to extracting the electroweak observable.
The experiment~\cite{WooBenCho97} determined the
ratio  $E_\mathrm{PNC}/\beta=1.5935(56) \, \mathrm{mV/cm}$, $\beta$ being the vector transition polarizability.
The most accurate value of $\beta$ comes from a combined
determination~\cite{BenWie99,DzuFlaGin02}, $\beta=-26.957(51) a_B^3$. With this $\beta$, we arrive at the nuclear weak charge
\begin{equation}
 Q_W(^{133}\mathrm{Cs}) = -73.16(29)_\mathrm{exp}(20)_\mathrm{th} \, , \label{Eq:QwResult}
\end{equation}
where the first uncertainty is experimental and the second uncertainty is theoretical.
Taking a weighted average, $\beta=-26.99(5) a_B^3$, of two determinations~\cite{BenWie99,VasSavSaf02}
results in $Q_W(^{133}\mathrm{Cs}) = -73.25(29)_\mathrm{exp}(20)_\mathrm{th}$. Both values are in a perfect agreement
with the prediction of the SM, $Q_W^\mathrm{SM} =-73.16(3)$ of Ref.~\cite{PDG08}.

Our result plays a unique, and at the same time complementary, role to collider experiments.
For $^{133}$Cs atom the relevant momentum transfer is just $\sim$ 30 MeV~\cite{MilSusTer02}, but the exquisite accuracy of
the interpretation probes minute contributions of the sea of virtual (including so-far undiscovered) particles
at a much higher mass scale. The new physics brought by the virtual sea is phenomenologically
described by weak isospin-conserving $S$ and isospin-breaking $T$ parameters~\cite{Ros02}: $\Delta Q_W=Q_W -Q_W^\mathrm{SM} = -0.800\, S - 0.007 T$.
At the $1 \sigma$-level, our result implies $|S|<0.45$. Parameter $S$ is important, for example, in indirectly constraining the
mass of the Higgs particle~\cite{Ros02}.
Similarly, the extra $Z$ boson, $Z'_\chi$, discussed in the introduction, would lead to a deviation~\cite{MarRos90}
$\Delta Q_W \approx 84 (M_W/M_{Z'_\chi})^2$, where $M_W$ is the mass of the $W$ boson.  Our result implies  50\% chance that  there is $Z'$ (i.e., $\Delta Q_W>0$). We find (at 84\% confidence level, including $M_{Z'_\chi}=\infty$) $M_{Z'_\chi} > 1.3 \, \mathrm{TeV}/c^2$, raising
the present lower bound on  the $Z'_\chi$ mass from direct collider searches~\cite{AalAbuAde07}. Our raised bound on the $Z'$ mass carves out a lower-energy part of the discovery reach of the LHC.

Our result confirms fundamental ``running'' (energy-dependence) of the electroweak force~\cite{CzaMar05,ErlRam05}.
The interaction strength of particles depends on their relative collision energy $E$: at higher energies the collision partners tend to penetrate deeper inside the shielding clouds of virtual particles surrounding the particles.
According to the SM, the interaction strength at low energies differs by about 3\% from its measured value at 100 GeV.
Compared to collider experiments, our result provides a reference point for the least energetic electroweak interactions.
Notice that the previous analyses~\cite{ShaPacTup05,DerPor07} of  PNC in Cs were consistent with no running~\cite{E158-PhysToday}.
With our $Q_W$, we find the effective interaction strength (we use scheme of Ref.~\cite{CzaMar00}),
 $\sin^2\!\theta_W^\mathrm{eff}(E \rightarrow 0)$ = 0.2381(11), $\theta_W$ being the Weinberg angle. The uncertainty is somewhat
 better than that of the previous most precise  low-energy test of the electroweak sector obtained
 in the electron scattering experiment at SLAC~\cite{E158collab}.
Our result is in agreement with the SM value~\cite{CzaMar00} of 0.2381(6).
While an earlier evidence for running of $\sin^2\!\theta_W$ has been obtained at SLAC~\cite{E158collab}, the prediction of the SM was outside of their  error bars.
In this regard, in addition to placing important constraints on new physics beyond the SM, our work provides a higher-confidence confirmation of the predicted running of the electroweak coupling at low energies.
In combination with the results of high-energy  experiments at SLAC and CERN~\cite{PDG08},
our work confirms the predicted running of the electroweak interaction over an  energy range spanning four
orders of magnitude (from $\sim 10$ MeV  to  $\sim 100$ GeV).


We thank O. Sushkov, M. Kozlov, J. Erler, W. Marciano, and M. Ramsey-Mussolf for discussions.
This work was initiated with support from the NIST precision measurement grant program and
supported in part by the NSF. S.G.P. was additionally supported by the RFBR under
grants No.~07-02-00210-a and 08-02-00460-a.


\begin{thebibliography}{32}
\expandafter\ifx\csname natexlab\endcsname\relax\def\natexlab#1{#1}\fi
\expandafter\ifx\csname bibnamefont\endcsname\relax
  \def\bibnamefont#1{#1}\fi
\expandafter\ifx\csname bibfnamefont\endcsname\relax
  \def\bibfnamefont#1{#1}\fi
\expandafter\ifx\csname citenamefont\endcsname\relax
  \def\citenamefont#1{#1}\fi
\expandafter\ifx\csname url\endcsname\relax
  \def\url#1{\texttt{#1}}\fi
\expandafter\ifx\csname urlprefix\endcsname\relax\def\urlprefix{URL }\fi
\providecommand{\bibinfo}[2]{#2}
\providecommand{\eprint}[2][]{\url{#2}}

\bibitem[{\citenamefont{Marciano and Rosner}(1990)}]{MarRos90}
\bibinfo{author}{\bibfnamefont{W.~J.} \bibnamefont{Marciano}} \bibnamefont{and}
  \bibinfo{author}{\bibfnamefont{J.~L.} \bibnamefont{Rosner}},
  \bibinfo{journal}{Phys.\ Rev.\ Lett.} \textbf{\bibinfo{volume}{65}},
  \bibinfo{pages}{2963} (\bibinfo{year}{1990}), \bibinfo{note}{{\bf 68}, 898(E)
  (1992)}.

\bibitem[{\citenamefont{Ramsey-Musolf}(1999)}]{Ram99}
\bibinfo{author}{\bibfnamefont{M.~J.} \bibnamefont{Ramsey-Musolf}},
  \bibinfo{journal}{Phys.\ Rev.\ C} \textbf{\bibinfo{volume}{60}},
  \bibinfo{pages}{015501/1} (\bibinfo{year}{1999}).

\bibitem[{\citenamefont{Wood et~al.}(1997)\citenamefont{Wood, Bennett, Cho,
  Masterson, Roberts, Tanner, and Wieman}}]{WooBenCho97}
\bibinfo{author}{\bibfnamefont{C.~S.} \bibnamefont{Wood}},
  \bibinfo{author}{\bibfnamefont{S.~C.} \bibnamefont{Bennett}},
  \bibinfo{author}{\bibfnamefont{D.}~\bibnamefont{Cho}},
  \bibinfo{author}{\bibfnamefont{B.~P.} \bibnamefont{Masterson}},
  \bibinfo{author}{\bibfnamefont{J.~L.} \bibnamefont{Roberts}},
  \bibinfo{author}{\bibfnamefont{C.~E.} \bibnamefont{Tanner}},
  \bibnamefont{and} \bibinfo{author}{\bibfnamefont{C.~E.}
  \bibnamefont{Wieman}}, \bibinfo{journal}{Science}
  \textbf{\bibinfo{volume}{275}}, \bibinfo{pages}{1759} (\bibinfo{year}{1997}).

\bibitem[{\citenamefont{Bennett and Wieman}(1999)}]{BenWie99}
\bibinfo{author}{\bibfnamefont{S.~C.} \bibnamefont{Bennett}} \bibnamefont{and}
  \bibinfo{author}{\bibfnamefont{C.~E.} \bibnamefont{Wieman}},
  \bibinfo{journal}{Phys.\ Rev.\ Lett.} \textbf{\bibinfo{volume}{82}},
  \bibinfo{pages}{2484} (\bibinfo{year}{1999}).

\bibitem[{Zpr()}]{ZprimeRefs}
\bibinfo{note}{J. Erler and P. Langacker, Phys. Rev. Lett. {\bf 84}, 212
  (2000); T. G. Rizzo, arXiv:0610104; P. Langacker, arXiv.org:0801.1345}.

\bibitem[{\citenamefont{Aaltonen et~al.}(2007)}]{AalAbuAde07}
\bibinfo{author}{\bibfnamefont{T.}~\bibnamefont{Aaltonen}}
  \bibnamefont{et~al.}, \bibinfo{journal}{Phys. Rev. Lett.}
  \textbf{\bibinfo{volume}{99}}, \bibinfo{pages}{171802}
  (\bibinfo{year}{2007}).

\bibitem[{\citenamefont{Khriplovich}(1991)}]{Khr91}
\bibinfo{author}{\bibfnamefont{I.~B.} \bibnamefont{Khriplovich}},
  \emph{\bibinfo{title}{Parity non-conservation in atomic phenomena}}
  (\bibinfo{publisher}{Gordon and Breach}, \bibinfo{address}{New York},
  \bibinfo{year}{1991}).

\bibitem[{\citenamefont{Bouchiat and Bouchiat}(1997)}]{BouBou97}
\bibinfo{author}{\bibfnamefont{M.-A.} \bibnamefont{Bouchiat}} \bibnamefont{and}
  \bibinfo{author}{\bibfnamefont{C.}~\bibnamefont{Bouchiat}},
  \bibinfo{journal}{Rep.\, Prog.\, Phys.} \textbf{\bibinfo{volume}{60}},
  \bibinfo{pages}{1351} (\bibinfo{year}{1997}).

\bibitem[{\citenamefont{Ginges and Flambaum}(2004)}]{GinFla04}
\bibinfo{author}{\bibfnamefont{J.~S.~M.} \bibnamefont{Ginges}}
  \bibnamefont{and} \bibinfo{author}{\bibfnamefont{V.~V.}
  \bibnamefont{Flambaum}}, \bibinfo{journal}{Phys. Rep.}
  \textbf{\bibinfo{volume}{397}}, \bibinfo{pages}{63} (\bibinfo{year}{2004}).

\bibitem[{\citenamefont{Dzuba et~al.}(1989)\citenamefont{Dzuba, Flambaum, and
  Sushkov}}]{DzuFlaSus89}
\bibinfo{author}{\bibfnamefont{V.}~\bibnamefont{Dzuba}},
  \bibinfo{author}{\bibfnamefont{V.~V.} \bibnamefont{Flambaum}},
  \bibnamefont{and} \bibinfo{author}{\bibfnamefont{O.}~\bibnamefont{Sushkov}},
  \bibinfo{journal}{Phys. Lett. A} \textbf{\bibinfo{volume}{140}},
  \bibinfo{pages}{493} (\bibinfo{year}{1989}).

\bibitem[{\citenamefont{Blundell et~al.}(1990)\citenamefont{Blundell, Johnson,
  and Sapirstein}}]{BluJohSap90}
\bibinfo{author}{\bibfnamefont{S.~A.} \bibnamefont{Blundell}},
  \bibinfo{author}{\bibfnamefont{W.~R.} \bibnamefont{Johnson}},
  \bibnamefont{and}
  \bibinfo{author}{\bibfnamefont{J.}~\bibnamefont{Sapirstein}},
  \bibinfo{journal}{Phys.\ Rev.\ Lett.} \textbf{\bibinfo{volume}{65}},
  \bibinfo{pages}{1411} (\bibinfo{year}{1990}).

\bibitem[{\citenamefont{Derevianko}(2000)}]{Der00}
\bibinfo{author}{\bibfnamefont{A.}~\bibnamefont{Derevianko}},
  \bibinfo{journal}{Phys.\ Rev.\ Lett.} \textbf{\bibinfo{volume}{85}},
  \bibinfo{pages}{1618} (\bibinfo{year}{2000}).

\bibitem[{\citenamefont{Derevianko}(2001)}]{Der02}
\bibinfo{author}{\bibfnamefont{A.}~\bibnamefont{Derevianko}},
  \bibinfo{journal}{Phys.\ Rev.\ A} \textbf{\bibinfo{volume}{65}},
  \bibinfo{pages}{012106} (\bibinfo{year}{2001}).

\bibitem[{the()}]{theorPNCextraNewBefore2005}
\bibinfo{note}{O. P. Sushkov, Phys. Rev. A {\bf 63}, 042504 (2001); V. A. Dzuba
  {\it et al.}, Phys. Rev. A {\bf 63}, 044103 (2001); M. G. Kozlov {\it et
  al.}, Phys. Rev. Lett. {\bf 86}, 3260 (2001); W. R. Johnson {\it et al.},
  Phys. Rev. Lett. {\bf 87}, 233001 (2001); M. Yu.\ Kuchiev and V. Flambaum,
  Phys.\ Rev.\ Lett. {\bf 89}, 283002 (2002); J. Sapirstein {\it et al.},
  Phys.\ Rev.\ A {\bf 67}, 052110 (2003); A.I. Milstein {\it et al.}, Phys.\
  Rev.\ A {\bf 67}, 62103 (2003)}.

\bibitem[{\citenamefont{Milstein et~al.}(2002)\citenamefont{Milstein, Sushkov,
  and Terekhov}}]{MilSusTer02}
\bibinfo{author}{\bibfnamefont{A.~I.} \bibnamefont{Milstein}},
  \bibinfo{author}{\bibfnamefont{O.~P.} \bibnamefont{Sushkov}},
  \bibnamefont{and} \bibinfo{author}{\bibfnamefont{I.~S.}
  \bibnamefont{Terekhov}}, \bibinfo{journal}{Phys.\ Rev.\ Lett.}
  \textbf{\bibinfo{volume}{89}}, \bibinfo{pages}{283003}
  (\bibinfo{year}{2002}).

\bibitem[{\citenamefont{Shabaev et~al.}(2005)\citenamefont{Shabaev, Pachucki,
  Tupitsyn, and Yerokhin}}]{ShaPacTup05}
\bibinfo{author}{\bibfnamefont{V.~M.} \bibnamefont{Shabaev}},
  \bibinfo{author}{\bibfnamefont{K.}~\bibnamefont{Pachucki}},
  \bibinfo{author}{\bibfnamefont{I.~I.} \bibnamefont{Tupitsyn}},
  \bibnamefont{and} \bibinfo{author}{\bibfnamefont{V.~A.}
  \bibnamefont{Yerokhin}}, \bibinfo{journal}{Phys. Rev. Lett.}
  \textbf{\bibinfo{volume}{94}}, \bibinfo{eid}{213002} (\bibinfo{year}{2005}).

\bibitem[{\citenamefont{Derevianko and Porsev}(2007)}]{DerPor07}
\bibinfo{author}{\bibfnamefont{A.}~\bibnamefont{Derevianko}} \bibnamefont{and}
  \bibinfo{author}{\bibfnamefont{S.~G.} \bibnamefont{Porsev}},
  \bibinfo{journal}{Eur. Phys. J. A} \textbf{\bibinfo{volume}{32}},
  \bibinfo{pages}{517} (\bibinfo{year}{2007}).

\bibitem[{\citenamefont{Bouchiat and Piketty}(1986)}]{BouPik86}
\bibinfo{author}{\bibfnamefont{C.}~\bibnamefont{Bouchiat}} \bibnamefont{and}
  \bibinfo{author}{\bibfnamefont{C.~A.} \bibnamefont{Piketty}},
  \bibinfo{journal}{Europhys. Lett.} \textbf{\bibinfo{volume}{2}},
  \bibinfo{pages}{511} (\bibinfo{year}{1986}).

\bibitem[{\citenamefont{Lindgren and Morrison}(1986)}]{LinMor86}
\bibinfo{author}{\bibfnamefont{I.}~\bibnamefont{Lindgren}} \bibnamefont{and}
  \bibinfo{author}{\bibfnamefont{J.}~\bibnamefont{Morrison}},
  \emph{\bibinfo{title}{Atomic Many--Body Theory}}
  (\bibinfo{publisher}{Springer--Verlag}, \bibinfo{address}{Berlin},
  \bibinfo{year}{1986}), \bibinfo{edition}{2nd} ed.

\bibitem[{\citenamefont{Dzuba et~al.}(2002)\citenamefont{Dzuba, Flambaum, and
  Ginges}}]{DzuFlaGin02}
\bibinfo{author}{\bibfnamefont{V.}~\bibnamefont{Dzuba}},
  \bibinfo{author}{\bibfnamefont{V.}~\bibnamefont{Flambaum}}, \bibnamefont{and}
  \bibinfo{author}{\bibfnamefont{J.}~\bibnamefont{Ginges}},
  \bibinfo{journal}{Phys. Rev. D} \textbf{\bibinfo{volume}{66}},
  \bibinfo{pages}{076013} (\bibinfo{year}{2002}).

\bibitem[{\citenamefont{Derevianko and Porsev}(2005)}]{DerPor05}
\bibinfo{author}{\bibfnamefont{A.}~\bibnamefont{Derevianko}} \bibnamefont{and}
  \bibinfo{author}{\bibfnamefont{S.~G.} \bibnamefont{Porsev}},
  \bibinfo{journal}{Phys. Rev. A} \textbf{\bibinfo{volume}{71}},
  \bibinfo{eid}{032509} (\bibinfo{year}{2005}).

\bibitem[{\citenamefont{Porsev and Derevianko}(2006)}]{PorDer06Na}
\bibinfo{author}{\bibfnamefont{S.~G.} \bibnamefont{Porsev}} \bibnamefont{and}
  \bibinfo{author}{\bibfnamefont{A.}~\bibnamefont{Derevianko}},
  \bibinfo{journal}{Phys.\ Rev.\ A} \textbf{\bibinfo{volume}{73}},
  \bibinfo{eid}{012501} (\bibinfo{year}{2006}).

\bibitem[{\citenamefont{Derevianko et~al.}(2008)\citenamefont{Derevianko,
  Porsev, and Beloy}}]{DerPorBel08}
\bibinfo{author}{\bibfnamefont{A.}~\bibnamefont{Derevianko}},
  \bibinfo{author}{\bibfnamefont{S.~G.} \bibnamefont{Porsev}},
  \bibnamefont{and} \bibinfo{author}{\bibfnamefont{K.}~\bibnamefont{Beloy}},
  \bibinfo{journal}{Phys. Rev. A} \textbf{\bibinfo{volume}{78}},
  \bibinfo{eid}{010503} (\bibinfo{year}{2008}).

\bibitem[{lif()}]{lifetimesCsRefs}
\bibinfo{note}{L. Young {\it et al.}, Phys. Rev. A {\bf 50}, 2174 (1994); R. J.
  Rafac {\it et al.}, Phys. Rev. A {\bf 60}, 3648 (1999); A. Derevianko and S.
  G. Porsev, Phys. Rev. A {\bf 65}, 053403 (2002); M.-A. Bouchiat {\it et al.},
  J. Phys. (France) Lett. {\bf 45}, 523 (1984); S. C. Bennett {\it et al.},
  Phys. Rev. A {\bf 59}, R16 (1999)}.

\bibitem[{\citenamefont{Vasilyev et~al.}(2002)\citenamefont{Vasilyev, Savukov,
  Safronova, and Berry}}]{VasSavSaf02}
\bibinfo{author}{\bibfnamefont{A.~A.} \bibnamefont{Vasilyev}},
  \bibinfo{author}{\bibfnamefont{I.~M.} \bibnamefont{Savukov}},
  \bibinfo{author}{\bibfnamefont{M.~S.} \bibnamefont{Safronova}},
  \bibnamefont{and} \bibinfo{author}{\bibfnamefont{H.~G.} \bibnamefont{Berry}},
  \bibinfo{journal}{Phys. Rev. A} \textbf{\bibinfo{volume}{66}},
  \bibinfo{pages}{020101} (\bibinfo{year}{2002}).

\bibitem[{\citenamefont{Amsler et~al.}(2008)}]{PDG08}
\bibinfo{author}{\bibfnamefont{C.}~\bibnamefont{Amsler}} \bibnamefont{et~al.}
  (\bibinfo{collaboration}{Particle Data Group}), \bibinfo{journal}{Phys. Lett.
  B} \textbf{\bibinfo{volume}{667}}, \bibinfo{pages}{1} (\bibinfo{year}{2008}).

\bibitem[{\citenamefont{Rosner}(2002)}]{Ros02}
\bibinfo{author}{\bibfnamefont{J.~L.} \bibnamefont{Rosner}},
  \bibinfo{journal}{Phys. Rev. D} \textbf{\bibinfo{volume}{65}},
  \bibinfo{pages}{073026} (\bibinfo{year}{2002}).

\bibitem[{\citenamefont{Czarnecki and Marciano}({2005})}]{CzaMar05}
\bibinfo{author}{\bibfnamefont{A.}~\bibnamefont{Czarnecki}} \bibnamefont{and}
  \bibinfo{author}{\bibfnamefont{W.~J.} \bibnamefont{Marciano}},
  \bibinfo{journal}{Nature} \textbf{\bibinfo{volume}{{435}}},
  \bibinfo{pages}{437} (\bibinfo{year}{{2005}}).

\bibitem[{\citenamefont{Erler and Ramsey-Musolf}(2005)}]{ErlRam05}
\bibinfo{author}{\bibfnamefont{J.}~\bibnamefont{Erler}} \bibnamefont{and}
  \bibinfo{author}{\bibfnamefont{M.~J.} \bibnamefont{Ramsey-Musolf}},
  \bibinfo{journal}{Phys. Rev. D} \textbf{\bibinfo{volume}{72}},
  \bibinfo{eid}{073003} (\bibinfo{year}{2005}).

\bibitem[{\citenamefont{Schwarzschild}(2005)}]{E158-PhysToday}
\bibinfo{author}{\bibfnamefont{B.}~\bibnamefont{Schwarzschild}},
  \bibinfo{journal}{Phys. Today} \textbf{\bibinfo{volume}{58}},
  \bibinfo{pages}{23} (\bibinfo{year}{2005}).

\bibitem[{\citenamefont{Czarnecki and Marciano}({2000})}]{CzaMar00}
\bibinfo{author}{\bibfnamefont{A.}~\bibnamefont{Czarnecki}} \bibnamefont{and}
  \bibinfo{author}{\bibfnamefont{W.~J.} \bibnamefont{Marciano}},
  \bibinfo{journal}{Int. J. Mod. Phys. A} \textbf{\bibinfo{volume}{{15}}},
  \bibinfo{pages}{2365} (\bibinfo{year}{{2000}}).

\bibitem[{\citenamefont{Anthony et~al.}(2005)}]{E158collab}
\bibinfo{author}{\bibfnamefont{P.~L.} \bibnamefont{Anthony}}
  \bibnamefont{et~al.} (\bibinfo{collaboration}{SLAC E158}),
  \bibinfo{journal}{Phys. Rev. Lett.} \textbf{\bibinfo{volume}{95}},
  \bibinfo{pages}{081601} (\bibinfo{year}{2005}), \eprint{hep-ex/0504049}.

\end{thebibliography}

\end{document}